\documentclass{vospwks2007}

\usepackage{color}
\usepackage{epsf}

\title{GAVO Tools for the Analysis of Stars and Nebulae}
\author{Thomas Rauch}
\affil{Institut f\"ur Astronomie und Astrophysik, Eberhard Karls Universit\"at, T\"ubingen, Germany}

\begin{document}


\maketitle

\begin{abstract}
Within the framework of the German Astrophysical Virtual Observatory 
(GAVO),
we provide synthetic spectra, simulation software for the calculation of 
NLTE model atmospheres, as well as necessary atomic data. This will
enable a VO user to directly compare observation and model-atmosphere spectra
on three levels: The easiest and fastest way is the use of our pre-calculated 
flux-table grid in which one may inter- and extrapolate. For a more precise
analysis of an abservation, the VO user may improve the fit to the observation
by the calculation of individual model atmospheres with fine-tuned
photospheric parameters via the WWW interface {\sc TMAW}. The more
experienced VO user may create own atomic-data files for a more detailed
analysis and calculate model atmosphere and flux tables with these.
\end{abstract}

\section{Introduction}
\label{sect:introduction}
In the last two decades, we have developed 
{\sc TMAP\footnote{http://astro.uni-tuebingen.de/\raisebox{.2em}{{\tiny $\sim$}}rauch/TMAP/TMAP.html}}, 
the T\"ubingen NLTE Model-Atmosphere Package \citep{w1986, wea2003b,rd2003}.
{\sc TMAP} has been successfully employed to calculate fully line-blanketed model atmospheres for
hot, compact stars, e.g\@. \citet{rea2007}, under the assumptions of hydrostatic and radiative equilibrium,
plane-parallel geometry, and non-local thermodynamic equilibrium (NLTE).

In contrast to fully line-blanketed LTE simulations, 
such NLTE calculations are still a domain
of specialists. Our previous analyses concentrated on pre-white dwarfs like, e.g.,
central stars of planetary nebulae but our
simulations can be easily utilized to white dwarfs, hot subdwarfs, neutron stars, and
accretion disks in cataclysmic variables or X-ray binaries.
We have presently arrived at a high level of sophistication and include opacities of all elements
from hydrogen to nickel.

For other simulation software which might be included in GAVO in future, 
the experience with {\sc TMAP} within this project will help to extend the VO spectral analysis 
to other stellar spectral types.

\section{Spectral Analysis via GAVO}
\label{sect:spectralanalysis}

We aim to provide synthetic spectra in order to compare these directly to observed spectra within the VO.
In the case of stellar spectra (Sect.\,\ref{subsect:stars}), flux tables will be provided from the X-ray to the
infrared wavelength range (Fig.\,\ref{fig:TMAW}).

\subsection{Stars}
\label{subsect:stars}

Based on {\sc TMAP} (Sect\@. \ref{sect:introduction}), GAVO aims to provide
(please note that the given URLs will change to the GAVO portal\footnote{http://www.g-vo.org/portal/} later)
\vspace{5mm}\\
--  Synthetic Spectra    \hspace{ 5.1mm}({\sc TMAF}\footnote{http://astro.uni-tuebingen.de/\raisebox{.2em}{\tiny $\sim$}rauch/TMAF/TMAF.html})\vspace{3mm}\\
--  Simulation Software  \hspace{ 1.0mm}({\sc TMAW}\footnote{http://astro.uni-tuebingen.de/\raisebox{.2em}{\tiny $\sim$}rauch/TMAW/TMAW.shtml})\vspace{3mm}\\
--  Atomic Data          \hspace{12.0mm}({\sc TMAD}\footnote{http://astro.uni-tuebingen.de/\raisebox{.2em}{\tiny $\sim$}rauch/TMAD/TMAD.html})\vspace{5mm}\\

With this offer, the way to a reliable spectral analysis for the VO user is threefold:

\begin{figure}[ht]
\epsfxsize=8cm
\epsffile{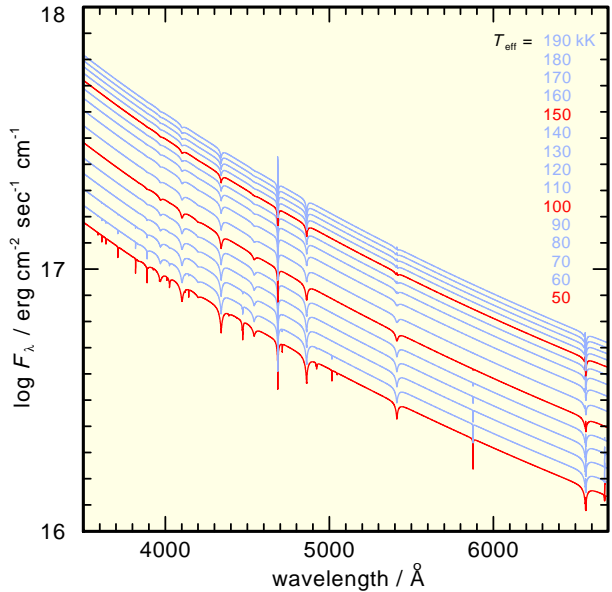}
\caption{Example of our pre-calculated flux-table grids 
         ($\log g = 7$, $\{X_{\mathrm{H}}\} = \{X_{\mathrm{He}}\} = 0.5$)
         which are available via TMAF.}
\label{fig:H+He}
\end{figure}

\begin{itemize}
\item
   A fast and easy spectral analysis is provided for
   an observation-orientated VO user who may analyze the performed observations with 
   existing model-flux grids ({\sc TMAF}) which are ready to use and 
   well-suited for inter- or extrapolation (within limits).
   The grids span generally over a wide range of effective temperature 
   ($T_{\mathrm{eff}} = 50 - 190\,\mathrm{kK}$) 
   and surface gravity 
   ($\log g = 5 - 9$)
   for different chemical compositions, 
   e.g., pure H, pure He, H+He (Fig\@. \ref{fig:H+He}), He+C+N+O, 
   H -- Ca \citep{r1997}, and H -- Ni \citep{r2003}.
   \vspace{3mm}
\item
   For more detailed investigations of a specific object, where the use of pre-calculated model-grid fluxes
   is not sufficient, the VO user may calculate individual 
   model atmospheres based on standard model atoms -- neither profound knowledge of theory nor
   experience with the software is here a pre-requisite. The photospheric parameters
   $T_{\mathrm{eff}}$, $\log g$, and mass fractions $\{X_{\mathrm{i}}\}$
   for $i \in \left[\mathrm{H, He, C, N, O}\right]$ can be adjusted in order to improve the fit to the observation.
   This is performed via {\sc TMAW}, a WWW service within GAVO.
   \vspace{3mm}
\item
   For more experienced observers and theoreticians, who want to compare e.g\@. their own simulations with
   results of {\sc TMAP}, the creation and upload of own atomic-data files is possible.
   We will provide model atoms which are suited for the use by {\sc TMAP}. These may be adjusted for an
   individual object.
\end{itemize}
\vspace{10mm}

\begin{figure}[ht]
\epsfxsize=8cm
\epsffile{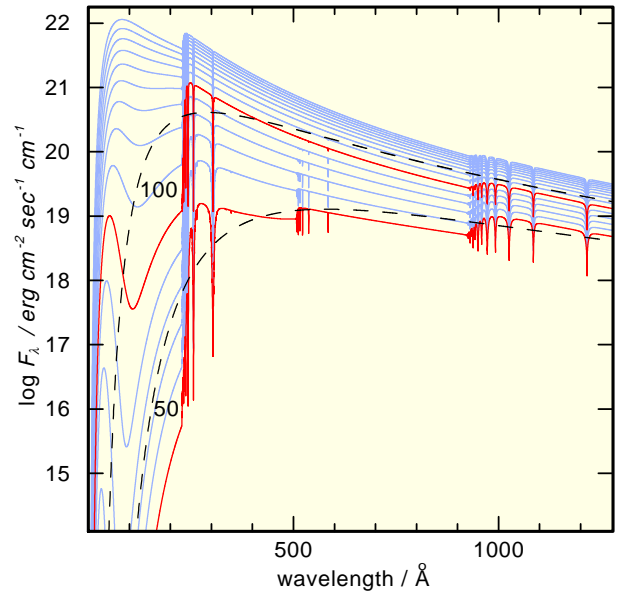}
\caption{Flux tables calculated from the same models as shown in Fig\@. \ref{fig:H+He}.
         For $T_{\mathrm{eff}} =$ 50\,kK and 100\,kK, blackbody flux distrubutions are shown
         in order to demonstrate the deviations.}
\label{fig:H+He_ion}
\end{figure}

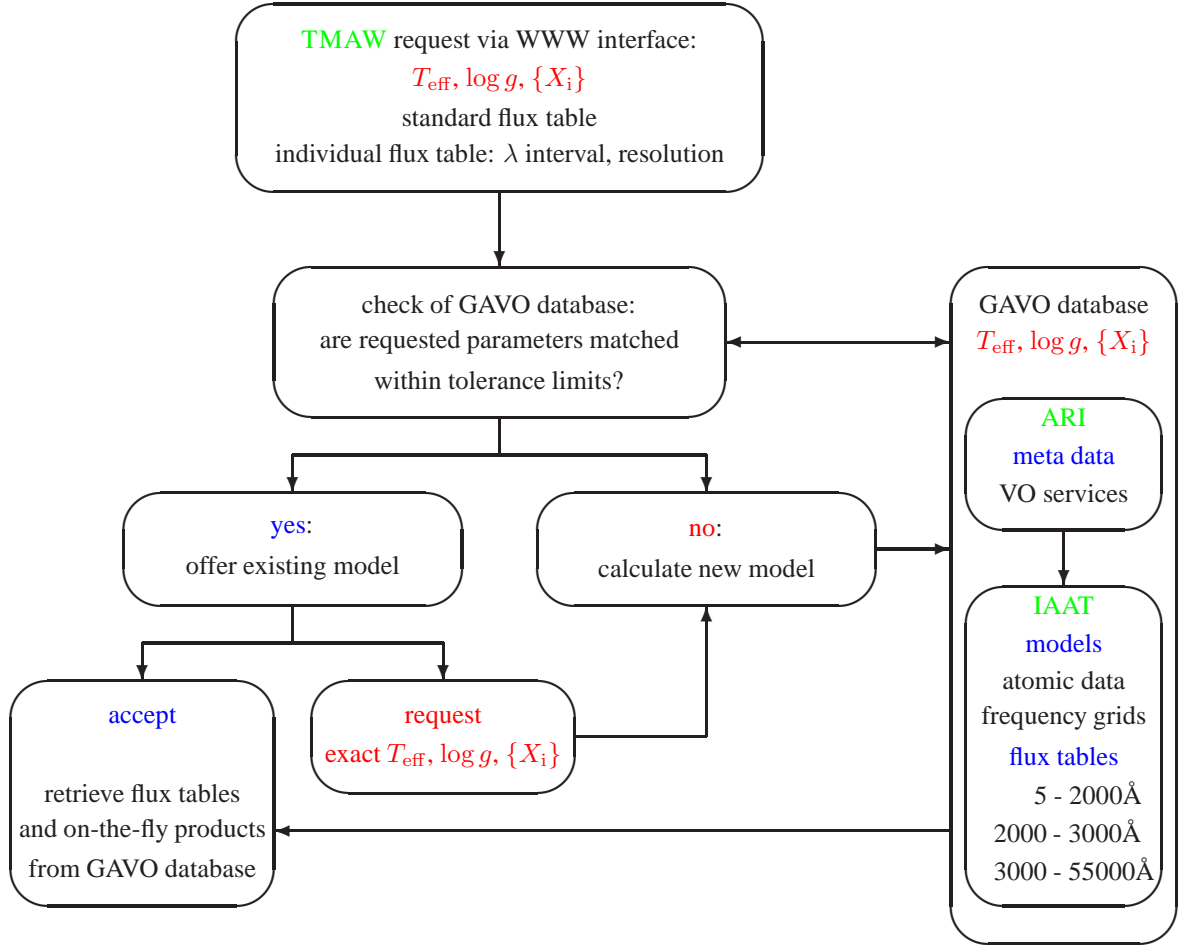
\begin{figure*}[ht]
\begin{picture}(17.0,17.5)
\thicklines
%
%
%
\put( 4.5,17.0){\oval( 1.0, 1.0)[tl]}
\put(10.5,17.0){\oval( 1.0, 1.0)[tr]}
\put( 4.5,15.5){\oval( 1.0, 1.0)[bl]}
\put(10.5,15.5){\oval( 1.0, 1.0)[br]}
\put( 4.5,17.5){\line( 1, 0){ 6.0}}
\put( 4.5,15.0){\line( 1, 0){ 6.0}}
\put( 4.0,15.5){\line( 0, 1){ 1.5}}
\put(11.0,15.5){\line( 0, 1){ 1.5}}
\put( 4.0,16.5){\makebox( 7.0, 1.0)[c]{
{{\color{green}\sc TMAW} request via WWW interface:}
}}
\put( 4.0,16.0){\makebox( 7.0, 1.0)[c]{
{\color{red}$T_{\mathrm{eff}}$, $\log g$, $\{X_{\mathrm{i}}\}$}
}}
\put( 4.0,15.5){\makebox( 7.0, 1.0)[c]{
{standard flux table}
}}
\put( 4.0,15.0){\makebox( 7.0, 1.0)[c]{
{individual flux table: $\lambda$ interval, resolution}
}}
\put( 7.5,15.0){\vector( 0,-1){ 1.0}}
\put( 5.0,13.5){\oval( 1.0, 1.0)[tl]}
\put(10.0,13.5){\oval( 1.0, 1.0)[tr]}
\put( 5.0,12.5){\oval( 1.0, 1.0)[bl]}
\put(10.0,12.5){\oval( 1.0, 1.0)[br]}
\put( 5.0,14.0){\line( 1, 0){ 5.0}}
\put( 5.0,12.0){\line( 1, 0){ 5.0}}
\put( 4.5,12.5){\line( 0, 1){ 1.0}}
\put(10.5,12.5){\line( 0, 1){ 1.0}}
\put( 4.0,12.0){\makebox( 7.0, 3.0)[c]{
check of GAVO database:
}}
\put( 4.0,11.5){\makebox( 7.0, 3.0)[c]{
are requested parameters matched
}}
\put( 4.0,11.0){\makebox( 7.0, 3.0)[c]{
within tolerance limits?
}}
\put(10.5,13.0){\vector( 1, 0){ 3.0}}
\put(13.5,13.0){\vector(-1, 0){ 3.0}}
\put( 7.50,12.00){\line( 0,-1){ 0.5}}
\put( 4.75,11.50){\line( 1, 0){ 5.5}}
\put( 4.75,11.50){\vector( 0,-1){ 0.5}}
\put(10.25,11.50){\vector( 0,-1){ 0.5}}
%
%
%
\put( 3.0,10.5){\oval( 1.0, 1.0)[tl]}
\put( 6.5,10.5){\oval( 1.0, 1.0)[tr]}
\put( 3.0,10.0){\oval( 1.0, 1.0)[bl]}
\put( 6.5,10.0){\oval( 1.0, 1.0)[br]}
\put( 3.0,11.0){\line( 1, 0){ 3.5}}
\put( 3.0, 9.5){\line( 1, 0){ 3.5}}
\put( 2.5,10.0){\line( 0, 1){ 0.5}}
\put( 7.0,10.0){\line( 0, 1){ 0.5}}
\put( 2.0,10.0){\makebox( 5.5, 1.0)[c]{
{\color{blue}yes}:
}}
\put( 2.0, 9.5){\makebox( 5.5, 1.0)[c]{
offer existing model
}}
%
%
%
\put( 8.5,10.5){\oval( 1.0, 1.0)[tl]}
\put(12.0,10.5){\oval( 1.0, 1.0)[tr]}
\put( 8.5,10.0){\oval( 1.0, 1.0)[bl]}
\put(12.0,10.0){\oval( 1.0, 1.0)[br]}
\put( 8.5,11.0){\line( 1, 0){ 3.5}}
\put( 8.5, 9.5){\line( 1, 0){ 3.5}}
\put( 8.0,10.0){\line( 0, 1){ 0.5}}
\put(12.5,10.0){\line( 0, 1){ 0.5}}
\put( 7.5,10.0){\makebox( 5.5, 1.0)[c]{
{\color{red}no}:
}}
\put( 7.5, 9.5){\makebox( 5.5, 1.0)[c]{
calculate new model
}}
\put(12.5,10.25){\vector( 1, 0){ 1.0}}
\put( 4.75, 9.50){\line( 0,-1){ 0.5}}
\put( 2.75, 9.00){\line( 1, 0){ 4.0}}
\put( 2.75, 9.00){\vector( 0,-1){ 0.5}}
\put( 6.75, 9.00){\vector( 0,-1){ 0.5}}
%
%
%
\put( 1.5, 8.0){\oval( 1.0, 1.0)[tl]}
\put( 4.0, 8.0){\oval( 1.0, 1.0)[tr]}
\put( 1.5, 6.0){\oval( 1.0, 1.0)[bl]}
\put( 4.0, 6.0){\oval( 1.0, 1.0)[br]}
\put( 1.5, 8.5){\line( 1, 0){ 2.5}}
\put( 1.5, 5.5){\line( 1, 0){ 2.5}}
\put( 1.0, 6.0){\line( 0, 1){ 2.0}}
\put( 4.5, 6.0){\line( 0, 1){ 2.0}}
\put( 0.5, 7.5){\makebox( 4.5, 1.0)[c]{
{\color{blue}accept}
}}
\put( 0.5, 6.5){\makebox( 4.5, 1.0)[c]{
{retrieve flux tables}
}}
\put( 0.5, 6.0){\makebox( 4.5, 1.0)[c]{
{and on-the-fly products}
}}
\put( 0.5, 5.5){\makebox( 4.5, 1.0)[c]{
{from GAVO database}
}}
\put(13.50, 6.50){\vector(-1,0){ 9.0}}
%
%
%
\put( 5.5, 8.0){\oval( 1.0, 1.0)[tl]}
\put( 8.0, 8.0){\oval( 1.0, 1.0)[tr]}
\put( 5.5, 7.5){\oval( 1.0, 1.0)[bl]}
\put( 8.0, 7.5){\oval( 1.0, 1.0)[br]}
\put( 5.5, 8.5){\line( 1, 0){ 2.5}}
\put( 5.5, 7.0){\line( 1, 0){ 2.5}}
\put( 5.0, 7.5){\line( 0, 1){ 0.5}}
\put( 8.5, 7.5){\line( 0, 1){ 0.5}}
\put( 4.5, 7.5){\makebox( 4.5, 1.0)[c]{
{\color{red}request}
}}
\put( 4.5, 7.0){\makebox( 4.5, 1.0)[c]{
{\color{red} exact $T_{\mathrm{eff}}$, $\log g$, $\{X_{\mathrm{i}}\}$}
}}
\put( 8.50, 7.75){\line( 1, 0){ 1.75}}
\put(10.25, 7.75){\vector( 0, 1){ 1.75}}
%
%
%
\put(14.0,13.5){\oval( 1.0, 1.0)[tl]}
\put(16.0,13.5){\oval( 1.0, 1.0)[tr]}
\put(14.0, 5.5){\oval( 1.0, 1.0)[bl]}
\put(16.0, 5.5){\oval( 1.0, 1.0)[br]}
\put(14.0,14.0){\line( 1, 0){ 2.0}}
\put(14.0, 5.0){\line( 1, 0){ 2.0}}
\put(13.5, 5.5){\line( 0, 1){ 8.0}}
\put(16.5, 5.5){\line( 0, 1){ 8.0}}
\put(13.5,12.0){\makebox( 3.0, 3.0)[c]{
GAVO database
}}
\put(13.5,11.5){\makebox( 3.0, 3.0)[c]{
{\color{red}$T_{\mathrm{eff}}$, $\log g$, $\{X_{\mathrm{i}}\}$}
}}
\put(14.2,11.75){\oval( 1.0, 1.0)[tl]}
\put(15.8,11.75){\oval( 1.0, 1.0)[tr]}
\put(14.2,11.00){\oval( 1.0, 1.0)[bl]}
\put(15.8,11.00){\oval( 1.0, 1.0)[br]}
\put(14.2,12.25){\line( 1, 0){ 1.6}}
\put(14.2,10.50){\line( 1, 0){ 1.6}}
\put(13.7,11.00){\line( 0, 1){ 0.75}}
\put(16.3,11.00){\line( 0, 1){ 0.75}}
\put(13.5,10.5){\makebox( 3.0, 3.0)[c]{
{\color{green}ARI}
}}
\put(13.5,10.0){\makebox( 3.0, 3.0)[c]{
{\color{blue}meta data}
}}
\put(13.5, 9.5){\makebox( 3.0, 3.0)[c]{
{VO services}
}}
\put(15.0,10.50){\vector( 0,-1){ 0.75}}
\put(14.2, 9.25){\oval( 1.0, 1.0)[tl]}
\put(15.8, 9.25){\oval( 1.0, 1.0)[tr]}
\put(14.2, 6.00){\oval( 1.0, 1.0)[bl]}
\put(15.8, 6.00){\oval( 1.0, 1.0)[br]}
\put(14.2, 9.75){\line( 1, 0){ 1.6}}
\put(14.2, 5.50){\line( 1, 0){ 1.6}}
\put(13.7, 6.00){\line( 0, 1){ 3.25}}
\put(16.3, 6.00){\line( 0, 1){ 3.25}}
\put(13.5, 8.0){\makebox( 3.0, 3.0)[c]{
{\color{green}IAAT}
}}
\put(13.5, 7.5){\makebox( 3.0, 3.0)[c]{
{\color{blue}models}
}}
\put(13.5, 7.0){\makebox( 3.0, 3.0)[c]{
{atomic data}
}}
\put(13.5, 6.5){\makebox( 3.0, 3.0)[c]{
{frequency grids}
}}
\put(13.5, 6.0){\makebox( 3.0, 3.0)[c]{
{\color{blue}flux tables}
}}
\put(13.5, 5.5){\makebox( 3.0, 3.0)[c]{
\hbox{}\hspace{5.5mm}5 - 2000\AA}}
\put(13.5, 5.0){\makebox( 3.0, 3.0)[c]{
2000 - 3000\AA}}
\put(13.5, 4.5){\makebox( 3.0, 3.0)[c]{
\hbox{}\hspace{1.8mm}3000 - 55000\AA
}}
\end{picture}\vspace{-40mm}
\caption{Scheme of TMAW. The VO user sends a flux-table request to the GAVO database by entering the
photospheric parameters in TMAW. If a suitable model is available within tolerance
limits, this is offered to the VO user. In case that the parameters are not extactly matched, the VO user
may decide to request a model with the exact parameters. TMAW will start a model-atmosphere calculation at the
IAAT then. As soon as the model is converged, the VO user can retrieve the flux table and various
on-the-fly products from the GAVO database.}\vspace{8mm}
\label{fig:TMAW}
\end{figure*}

{\sc TMAW} is a WWW interface which provides both, 
WWW access to existing model-flux grids as well as the possibility to calculate models and fluxes
using individual parameters. 
A scheme of the data flow is shown in Fig.\,\ref{fig:TMAW}. 
The {\sc TMAW} user has to enter the photospheric parameters via the 
{\sc TMAW} interface (Sect.\,\ref{sect:spectralanalysis}).

If a suitable model is already available in the GAVO database at the 
ARI\footnote{Astronomisches Rechen-Institut, Heidelberg, Germany},
it is offered to the VO user. This is done using preset tolerance limits in order to
speed up this process and to avoid the calculation of models at unreasonable small grid steps. 
In case that the VO user accepts this model, only the requested individual flux table will be calculated
and then sent to the VO user together with the requested standard flux table.
If the search is negative, a complete new model will be calculated at the 
IAAT\footnote{Institut f\"ur Astronomie und Astrophysik, T\"ubingen, Germany}.

Depending on the requested $T_{\mathrm{eff}}$ and $\{X_{\mathrm{i}}\}$, an atomic-data file is created
using either predefined model atoms or a VO user-created atomic-data file which has to be uploaded before.

The standard calculation starts with the computation of a model in grey approximation which is followed by a
NLTE model-atmosphere calculation in a number of steps (first, a so-called ``continuum'' model is
calculated which considers no line opacities and subsequently, a ``line'' model is calculated
which accounts for line-blanketing in addition). Once the model is converged (relative corrections
in temperature, densities, occupation numbers less that $10^{-4}$ in all depth points), a VO 
user-requested flux table as well as the three standard flux tables (Fig\@.\ref{fig:TMAW}) are calculated.  

As soon as the model has been calculated, a standard output will be sent to the VO user by email.
It comprises flux tables ($\lambda$, $F_\lambda$, $F_{\lambda} /  F_{\lambda, \mathrm{cont}}$)
for a selected wavelength range and resolution, plus one of the three standard flux tables,
a plot of the flux table and the output from the last iteration of the {\sc TMAP} model-atmosphere
calculation.

Simultaneously, the meta data of the calculated model will be sent to the ARI database 
which therefore will be growing in time. 
The model atmosphere, the respective atomic data and frequency grid files, as well as the standard flux
tables remain in the database at the IAAT.

\subsection{Nebulae}
\label{subsect:nebulae}

For the analysis of ionized nebulae, a variety of photoionization codes exists. 
Although realistic model-atmosphere fluxes are used sometimes, it is still common to
use easy-to-calculate blackbody-flux distributions to simulate the exciting star. 
Examples for deviations are shown, e.g., in \citet{r1997}, \citet{aea2003}, \citet{of2006},
and Fig\@. \ref{fig:H+He_ion}.

One of the standard flux tables ($5-2000\,\mathrm{\AA}$, binned to 0.1\,\AA\ intervals,
cf\@. Fig\@. \ref{fig:H+He_ion}) 
is well suited as ionizing spectrum for photoionization models of planetary nebulae. 
Such tables are already used, e.g., 
by {\sc CLOUDY} \citep{fea1998} and {\sc \mbox{MOCASSIN}} \citep{eea2005}. 
{\sc \mbox{MOCASSIN}} is already able to deal with {\sc TMAP} flux tables \citep[e.g\@.][]{eea2003} and consequently, 
we will set up a WWW interface for the control of {\sc \mbox{MOCASSIN}}  which makes directly use of 
model-atmosphere fluxes  within the GAVO database.
However, any photoionization code may benefit from the synthetic spectra provided by GAVO.

\section{Conclusions and future plans}
\label{sect:future}

Within this GAVO project, we will set up the basics to provide spectral analysis for the VO user.
The principal idea has to be that for a VO user this task should be as easy as to collect spectra 
of an object from the VO. Thus, we will provide spectral analysis for hot, compact stars with our 
{\sc TMAP} models at three levels.

The use of model-grid fluxes appears to be the easiest and fastest way for the VO user 
and it is unimportant how these have been calculated, i.e\@. no knowledge about the code 
is necessary. The calculation of model-grid fluxes with other codes like
{\sc PHOENIX} \citep{hb1999}, {\sc WRUNIQ} \citep{gea2003}, or WM-basic \citep{p2003},
which, e.g\@., account for mass loss and stellar winds could extend the database considerably. 

The easy use of other codes via WWW interfaces like {\sc TMAW} is highly desirable.
This work will be done by the respective working groups.

A more general problem is the use of uniformly formatted model atoms for the different existing
model-atmosphere codes like proposed for {\sc TMAD}
by \citet{rd2003}. This requires a concerted action by all modeling groups.

Precise spectral analysis requires extended grids of elaborated model atmospheres.
In the framework of GRID computing \citep{f2005, f2006}, 
the calculation of model-atmosphere grids and flux tables, e.g\@. via {\sc TMAW}, is an excellent 
application to efficiently calculate synthetic spectra on reasonable time scales.

\section*{Acknowledgments}
This work is supported by the \emph{German Astrophysical Virtual Observatory} project
of the German Federal Ministry of Education and Research (BMBF) under grant 05\,AC6VTB.
\vspace{7cm}
\hbox{}


\begin{thebibliography}{}


\bibitem[Armsdorfer et al\@.(2003)Armsdorfer, Kimeswenger, \& Rauch]{aea2003}
         Armsdorfer B., Kimeswenger S., \& Rauch T\@.
         2003,
         in: {\it Proc\@. IAU Symp\@. 209. Planetary Nebulae: Their Evolution and Role in the Universe},
         eds\@. S\@. Kwok, M\@. Dopita, R\@. Sutherland, p\@. 511
\bibitem[Ercolano et al\@.(2003)Ercolano et al\@.]{eea2003}
         Ercolano, B., Barlow, M\@. J., Storey, P\@. J., Liu, X.-W., Rauch, T., \& Werner, K\@.
         2003,
         MNRAS, 344, 1145
\bibitem[Ercolano et al\@.(2005)Ercolano et al\@.]{eea2005}
         Ercolano, B., Barlow, M\@. J., \& Storey, P\@. J\@.
         2005,
         \mbox{MNRAS}, 362, 1038
\bibitem[Ferland et al\@.(1998)Ferland et al\@.]{fea1998}
         Ferland, G\@. J., Korista, K\@. T., Verner, D\@. A., Ferguson, J\@. W., Kingdon, J\@. B., \& Verner, E\@. M\@. 
         1998, 
         PASP, 110, 761
\bibitem[Foster(2005)Foster]{f2005}
         Foster, I\@.
         2005,
         in: Proceedings of the IFIP International Conference on Network and Parallel Computing
         (NPC\,2005), 2
\bibitem[Foster(2006)Foster]{f2006}
         Foster, I\@.
         2006,
         in: The Virtual Observatory in Action: New Science, New Technology, and Next Generation Facilities, 
         26th meeting of the IAU, Special Session 3, 6
\bibitem[Gr\"afener et al\@.(2003)Gr\"afener, Koesterke, \& Hamann]{gea2003}
         Gr\"afener, G., Koesterke, L., \& Hamann, W.-R\@.
         2002, 
         A\&A, 387, 244
\bibitem[Hauschildt \& Baron(1999)Hauschildt \& Baron]{hb1999}
         Hauschildt, P\@. H., \& Baron, E\@.
         1999,
         Journal of Computational and Applied Mathematics, 109, 41
\bibitem[Osterbrock \& Ferland(2006)Osterbrock \& Ferland]{of2006}
         Osterbrock, D\@. E., \& Ferland, G\@. J\@.
         2006,
         Astrophysics of Gaseous Nebulae and Active Galactic Nuclei,
         2nd Edition,
         University Science Books, Sausalito, California
\bibitem[Pauldrach(2003)Pauldrach]{p2003}
         Pauldrach, A.W.A\@. 
         2003, 
         in: The Cosmic Circuit of Matter, 
         ed\@. E\@. Schielicke,
         Reviews in Modern Astronomy, Vol\@. 16, p\@. 133, New York: Wiley--VCH
\bibitem[Rauch(1997)Rauch]{r1997}
         Rauch, T\@.
         1997,
         A\&A, 320, 237
\bibitem[Rauch(2003)Rauch]{r2003}
         Rauch, T\@.
         2003,
         A\&A, 403, 709
\bibitem[Rauch \& Deetjen(2003)Rauch \& Deetjen]{rd2003}
         Rauch, T., \& Deetjen, J\@. L\@. 
         2003,
         in: Stellar Atmosphere Modeling,
         eds\@. I\@. Hubeny, D\@. Mihalas, K\@. Werner,
         The ASP Conference Series, Vol\@. 288, p\@. 103
\bibitem[Rauch et al\@.(2007)Rauch et al\@.]{rea2007}
         Rauch, T., Ziegler, M., Werner, K., Kruk, J\@. W., Oliveira, C\@. M., Vande Putte, D., Migniani, R\@. P., \& Kerber, F\@.
         2007,
         A\&A in press 
\bibitem[Werner(1986)Werner]{w1986}
         Werner, K\@.
         1986,
         A\&A, 161, 177
\bibitem[Werner et al\@.(2003)Werner et al\@.]{wea2003b}  
         Werner, K., Dreizler, S., Deetjen, J\@. L., et al\@. 
         2003,
         in: Stellar Atmosphere Modeling,
         eds\@. I\@. Hubeny, D\@. Mihalas, K\@. Werner,
         The ASP Conference Series, Vol\@. 288, p\@. 31

\end{thebibliography}
\end{document}